\documentclass[10pt,a4paper,twocolumn]{article}
\usepackage[top=30mm, bottom=30mm, left=15mm, right=15mm]{geometry}
\usepackage{amssymb,amsmath,latexsym}
\usepackage[font={small,it}]{caption}
\usepackage[T1]{fontenc}
\usepackage[utf8]{inputenc}
\usepackage{physics}
\usepackage{lmodern}
\usepackage{amsmath}
\usepackage{amssymb}
\usepackage{graphicx}
\usepackage{sectsty}
\usepackage{mathrsfs}
\usepackage{multirow}
\usepackage{mathtools}
\usepackage{tabularx}
\usepackage{array}
\usepackage{soul}
\usepackage{xcolor}
\usepackage[colorlinks=true,
            linkcolor=red,
            urlcolor=blue,
            citecolor=cyan]{hyperref}
\title{On the gravitational collapse in 4-dimensional Einstein-Gauss-Bonnet gravity}
\author{R. Hassannejad$^1$, A. Sadeghi$^1$ and F. Shojai$^{1}$ \\$^1$Department of Physics, University of Tehran, P.O. Box 14395-547\\Tehran, Iran.}
\begin{document}
\maketitle
\begin {abstract}
In this paper, we treat 4-dimensional Einstein-Gauss-Bonnet gravity as general relativity with an effective stress-energy tensor. We will study the modified Oppenheimer-Snyder-Datt model of the gravitational collapse of a star in a 4-dimensional Einstein-Gauss-Bonnet black hole.
The inside geometry of the star is described by the spatially flat Friedmann-Robertson-Walker metric and the matter is distributed uniformly without any pre-assumption about its equation of state. The exterior Einstein-Gauss-Bonnet black hole is smoothly matched to the interior geometry without the requirement of any thin shell. This gives the energy density, pressure, and the equation of state of collapsing matter. 
At the end, we study the time evolution of event and apparent horizons.
\end{abstract}
\section{Introduction}
The appearance of singularities in Einstein's theory of gravity is a well--known problem, but hard to deal with \cite{klsh}. Many studies acknowledge that a consistent and complete theory of quantum gravity will overcome this issue \cite{quantum}. Hence, it is usually believed that the regular black holes (BHs) geometries can effectively present the removal of singularity by quantum gravity effects. \cite{dhj,pnkl,lljo}.\\
There are two approaches to get rid of singularities inside the BHs. One can use Einstein's general relativity (GR) to
describe, for example, the evolution of a collapsing supermassive star while the quantum effects lead to violation of some energy conditions somewhere inside the star. Or one can assume that the stellar matter satisfies the energy conditions and there are some quantum modifications to GR acting as repulsive effects avoiding the singularity at the end of gravitational collapse.
This means that the modifications of Einstein's equations may be understood by incorporating the quantum effects either as
an effective energy-momentum tensor or as geometrical terms. Modifying Hilbert-Einstein action of GR by coupling it to the nonlinear electrodynamics by Bardeen \cite{lljo}, Sakharov's proposal $p=-\rho$ as the equation of state for the high-density region of a star \cite{sakh}, the quark matter as some kind of exotic matter at the core of neutron stars \cite{45-46Mal} and boson stars \cite{47-48Mal} follow the first approach. While the gravitational collapse of a thin light-like shell in a quantum gravity scenario \cite{34-35Mal}, the phase transition of quantum vacuum near the horizon in Gravastars \cite{25Mal} and black stars \cite{44Mal} follow the second one.\\
Following the first approach , recently, we have studied \cite{ourpaper} a collapsing star that evolves to a particular class of regular BHs according to the theory of GR. It is shown that assuming Oppenheimer-Snyder-Datt (OSD) model \cite{OSD},
the collapsing process is ended in an infinite time, therefore the singularity is avoided. For this purpose, the stellar matter should have uniform density and pressure and obey the polytropic equation of state. Also according to \cite{ourpaper}, when the star radius becomes smaller than a specific value, the stellar matter violates the strong energy condition (SEC). See \cite{ourpaper} for a detailed study of the time evolution of star surface, the event, and apparent horizons.\\
Recently a novel D-dimensional Einstein-Gauss-Bonnet (EGB) gravity was proposed \cite{plp}. The considered action consists of the usual Einstein-Hilbert term plus the Gauss-Bonnet term multiplied by a rescaled coupling constant, 
${\alpha}\to\frac{\alpha}{D-4}$. The regularized 4D EGB gravity admits singularity-free spherically symmetric metric in the sense that the gravitational potential approaches a finite value at short distances.
Through redefining the metric, the
authors of \cite{nilmk,ppgf,ppkk,nkb} have shown that one can reduce the 4D EGB theory to a scaler-tensor theory, which is a particular type of Horndeski's gravity. Therefore, it seems that there is a deep relation between the novel 4D EGB gravity and Horndeski's gravity.
The cosmological and spherically symmetric solutions of 4D EGB gravity are also investigated in \cite{hbjb,jnk} and \cite{vfsk,njj} respectively.
For instance, charged BHs in the presence of both linear and nonlinear electrodynamics and also rotating BHs along with their properties have been studied in \cite{Kumar,klkl,xzsrz}.
The quasinormal modes, the gravitational waves, and the shadows of 4D EGB BHs are investigated in \cite{pkjk,vfyj,xtcy,oorz,xffy,agji}. Moreover, some aspects of 4D EGB BH's have also been studied in \cite{pvdt,gfyu,spsjj,dncdkl,aeth,cfdt,jjkrz}, including thermodynamics and Hawking radiation. 
In the cosmological context, the modified Friedmann equations of such theory predict new exciting physics in the early universe \cite{njrf,nsxd} and reduce to the standard Friedmann equations at large scales. \\
On the other hand some authors pointed out the difficulties and defects of the regularized prescription of \cite{plp}. This motivated others to suggest new regularization methods and compare the results with \cite{plp}. However, it should be noted that the main result is that for spherically symmetric space-time, the BH solution of \cite{plp} is still valid even applying other regularization methods. 
It can be shown that in EGB approach \cite{plp}, there are divergent contributions to the field equations in four dimensions \cite{arre}. However according to \cite{jjkrz,aoki}, if the Weyl tensor of spatial metric of a D-dimensional solution of EGB gravity vanishes, the four dimensional limit of this solution is a solution of 4D EGB theory. The D-dimensional spherically symmetric solution of EGB gravity satisfies this condition \cite{jafarzade} and so its four dimensional limit is also a solution of 4D-EGB gravity.
In addition to Glavan and Lin regularization \cite{plp}, there are alternative regularizations  to construct other versions of the EGB theory in four-dimensions. This includes conformal regularization \cite{nilmk,fernandes1}, regularized Kaluza–Klein reduction \cite{ppkk,lu} which both lead to scalar–tensor theories. However, it is known that the most general scalar–tensor theory with second-order equations of motion and a conformally invariant scalar field can be reduced to the 4D-EGB theory \cite{plp} for some special values of coupling constants \cite{fer7}.
One of the consistent theories of 4D EGB gravity that only breaks the temporal diffeomorphism invariance, is proposed by \cite{ppkk, lu}. In the case of spherically symmetric metrics, this theory also gives the solution of Glavan and Lin \cite{plp}. An alternative method of regularization for 4D-EGB gravity is adding counter terms to the action such that the divergences of the action are removed. These methods of regularization are completely different from the other methods, however its static spherically symmetric black hole solutions are exactly the same as proposed in \cite{plp}.
In summary, the different regularized 4D-EGB theory also have the same static spherically symmetric BH solution as the original approach \cite{plp} (See \cite{fernandes1} and \cite{fer8} and references therein). Say in another way,  the approach of \cite{plp} works well in spherical symmetric geometry, and provides a well-defined action in this case \cite{klkl}. Therefore,  the black hole solution of \cite{plp} is still a valid solution and so it is worth to study the main features of it. \\
In \cite{naresh}, the gravitational collapse of a homogeneous dust sphere is discussed in 4D EGB gravity. It is shown that the presence of GB term leads to some delay in reaching the singularity  and moreover the collapse is ended with zero velocity. In this paper, we generalize the OSD collapsing model considering the novel 4D EGB gravity.
We study the formation of static spherically symmetric 4D EGB BH from the gravitational collapse of a spherical massive star without any pre-assumption about its matter content. For such a BH,  the curvature scalar diverges at short distances, while the gravitational potential approaches to a finite value and the gravitational force is repulsive. Therefore, it may be expected that an infalling particle never reaches the $r=0$ \cite{plp}. This is what we want to study in
detail here. \\
To do this, it is assumed that the star is described by the spatially flat Friedmann-Robertson-Walker (FRW) metric and the particles follow the radial geodesics at the star surface. To generalize the OSD collapsing model, the interior metric should be smoothly joined to the exterior 4D EGB BH metric.
In this way, we have obtained the density, pressure, the equation of state parameter of stellar matter. In addition, some geometrical properties of the star is obtained analytically. This includes the star surface, and event and apparent horizons. To find the properties of stellar matter, we consider the approach mentioned in the beginning of this section.  It is based on the Friedmann equations in which the GB term appears as an effective source.\\
The outline of this paper is as follows: In section \ref{sec2}, we recall EGB action and the 4-dimensional spherically symmetric metric. Then in section \ref{sec3}, we develop OSD gravitational collapse to the well-known BH solution of 4D EGB gravity. The equation of state of stellar matter is derived in section \ref{sec4}. Section \ref{sec5} deals with the evolution of the event and apparent horizons. We summarize our results in Section \ref{concluding}.\\
Throughout this paper, the signature of the metric tensor is assumed to be $(-, +, +, +)$. We use geometrized units, i.e., $G = c = 1$. A dimensionless variable is denoted with a tilde and obtained by normalizing it with
respect to the Schwarzschild radius, i.e. $\tilde{\tau} = \tau/2M$ and $\tilde{\alpha}=\alpha/(2M)^2$.
\section{4D spherically symmetric solution of EGB gravity}
\label{sec2}
The EGB action in D-dimensional space-time is written as \cite{plp,Kumar,kokk},
\begin{align}\label{A}
S_{D}= \int d^{D}x\sqrt{-g}\Big[\frac{R}{16\pi} & +\alpha\big(R_{\mu\nu\rho\sigma} R^{\mu\nu\rho\sigma}\notag\\&-4R_{\mu\nu}R^{\mu\nu}+R^2\big)\Big]
\end{align}
in which $\alpha$ is the GB coupling constant. The vacuum EGB field equations can be obtained via varying the action \eqref{A} with respect to $g_{\mu\nu}$ \cite{fer7}
\begin{align}\label{B}
G_{\mu\nu}+16\pi \alpha \mathcal{H}_{\mu\nu}=0
\end{align}
where $G_{\mu\nu}=R_{\mu\nu}-\frac{1}{2}g_{\mu\nu}R$ is the Einstein tensor and
\begin{align}\label{C}
\mathcal{H}_{\mu\nu}=&2(RR_{\mu\nu}-2R_{\mu\sigma}R^{\sigma}_{\nu}-2R_{\mu\sigma\nu\rho}R^{\sigma\rho}+R_{\mu\sigma\rho\delta}R^{\sigma\rho\delta}_{\nu})\notag\\
&-\frac{1}{2}(R_{\mu\nu\rho\sigma} R^{\mu\nu\rho\sigma}-4R_{\mu\nu}R^{\mu\nu}+R^2)g_{\mu\nu}
\end{align}
contains the higher-order curvature terms.
It should be remembered that, generally, in non-vacuum EGB gravity such as any modified theory of gravity, one can express the field equations in the Einstein form $G_{\mu\nu}=8\pi T_{\mu\nu}^{\text{eff}}$
where $T_{\mu\nu}^{\text{eff}}=T_{\mu\nu}^{\text{EGB}} +  T_{\mu\nu}^{\text{m}}$ is the effective stress-energy tensor which contains the matter stress-energy
tensor $T_{\mu\nu}^{\text{m}}$ and the curvature terms arising from EGB gravity, $T_{\mu\nu}^{\text{EGB}} =-2\alpha \mathcal{H}_{\mu\nu}$. This interpretation has been extensively used in $f(R)$ gravity \cite{f(r)},
curvature-matter couplings \cite{curvmatt},  Weyl gravity \cite{manheim} and in braneworlds \cite{sepangi}. In this approach, the effective stress energy tensor involves higher order curvature terms that is responsible for the violation of energy conditions. 
In $D=4$ the GB invariant term, $R_{\mu\nu\rho\sigma} R^{\mu\nu\rho\sigma}-4R_{\mu\nu}R^{\mu\nu}+R^2$ is a total derivative. Therefore, 
it does not contribute to the gravitational dynamics. \\
The spherically symmetric solution of (\ref{B}) and (\ref{C}) in D-dimension ($D\geq 5)$ is \cite{fer7,boulware,saha} 
\begin{align}
ds^2&=-f(r)dt^2+f^{-1}(r)dr^2+r^2d\Omega^2_{D-2},\nonumber\\
\label{E}
f(r)&=1+\frac{r^2}{32\pi(D-3)(D-4)\alpha}\Bigg[1\notag\\
&\pm\sqrt{{1+\frac{1024\pi^2(D-3)(D-4)\alpha M}{(D-2)\Omega_{D-2}r^{D-1}}}}\Bigg]
\end{align}
where $M$ and $\Omega_{D-2}$ are the mass parameter and the D-dimensional solid angle respectively. Redefining $\alpha\to\frac{\alpha}{D-4}$ in EGB action \eqref{A}, the line element \eqref{E} would be well-defined in $D=4$  \cite{plp}
\begin{align}\label{G}
f(r)=1+\frac{r^2}{32\pi\alpha}\Bigg[1\pm\sqrt{{1+\frac{128\pi\alpha M}{r^{3}}}}\Bigg]
\end{align} 
There are two branches of solutions for $\alpha>0$, and their asymptotic limit $r\to \infty$ are as follows,
\begin{align}\label{H}
f(r)_{-}=1-\frac{2M}{r},\hspace{0.5cm}f(r)_{+}=1+\frac{r^2}{16\pi\alpha}+\frac{2M}{r}
\end{align} 
which shows that the space-time is asymptotically AdS (flat) with negative (positive) gravitational mass for the $+(-)$ branch.
 In the limit of $r\rightarrow 0$ the metric \eqref{G} behaves as
\begin{align}\label{Har1}
f(r)_{-}=f(r)_{+}=1+ \sqrt{ \frac{r M}{8 \pi \alpha}}
\end{align} 
therefore it is non-singular at $r\rightarrow 0$.
In the following, we will focus on the $(-)$ branch of $\alpha>0$.\\
\section{OSD collapse to  4D EGB BH}
\label{sec3}
Now, we are going to investigate the OSD gravitational collapse to EGB BH \eqref{G} whose line element can be written as
\begin{align}
\label{I}
&ds^2 = - \Big(1-\frac{2M}{r\omega(r)}\Big)dt^2 + \Big(1-\frac{2M}{r\omega(r)}\Big)^{-1} dr^2 + r^2 d \Omega^2\nonumber \\
&\omega(r)=\frac{64\pi\alpha M}{r^3}\bigg[\Big(1+\frac{128\pi\alpha M}{r^3}\Big)^{\frac{1}{2}}-1\bigg]^{-1}
\end{align}
This will give a BH solution with two horizons if $M> \sqrt{16 \pi  \alpha}$. We assume that the stellar matter is described by a perfect fluid and the inside geometry is given by FRW metric which can be matched smoothly to the outside BH space-time \eqref{I}. This means that there is no thin shell distribution of matter.\\
In the conventional OSD collapsing model, the massive star is embedded in the Schwarzschild space-time. Therefore, the exterior region of the star is vacuum and it is smoothly matched to the interior FRW geometry, i.e. no energy layer appears on the star surface. Now consider a compact star embedded in the 4D EGB BH space-time (\ref{I}). We want to investigate under which conditions, there exists a smooth transition across the boundary of the star. This can be immediately obtained from the Israel junction conditions \cite{kojfg} expressed in appropriate coordinates. Writing the EGB metric \eqref{I} in terms of Painlev\'e-Gullstrand coordinates \cite{klpk}, one reads
\begin{align}\label{BB}
ds^2 = - d\tau^2 + (dr+ \sqrt{\frac{2M}{rw(r)}} d \tau)^2 + r^2 d\Omega^2
\end{align}
and the spatially flat FRW geometry is expressed as
\begin{align}\label{EE}
ds^2 = - d\tau^2 + (d r - r H(\tau) d \tau)^2 + r^2d\Omega^2
\end{align}
where $H( \tau)=\dot{ r}( \tau)/ r( \tau)$ is the correspoding Hubble parameter.\\
As mentioned before, here we have considered 4D EGB gravity as GR with an effective stress-energy tensor. Therefore, we use the junction conditions of GR. Otherwise, one should consider the EGB junction conditions \cite{dev}.
By considering \eqref{I}, it is straightforward to show that the components
of the extrinsic curvature on the star's surface are given by
\begin{align}\label{EWHH}
^{(in)}K^{\tau}_{\tau}&=0 \hspace{1.3cm} ^{(in)}K^{\theta}_{\theta}=^{(in)}K^{\phi}_{\phi}=\frac{1}{ R}\notag\\
^{(out)}K^{\tau}_{\tau}&=\frac{\dot{\gamma}}{ R}\hspace{1cm} ^{(out)}K^{\theta}_{\theta}=^{(out)}K^{\phi}_{\phi}=\frac{\gamma}{ R}
\end{align}
where $\gamma=\sqrt{\dot{ R}^2+1-2M/( Rw( R))}$. We find that for smooth transition at the boundary, $\gamma=1$. Therefore, in terms of dimensionless parameters, the surface of the star is given by
\begin{align}\label{CC}
\dot{R}(\tau) =-\frac{R}{4\sqrt{2 \pi\alpha}}\sqrt{(1+\frac{128 \pi\alpha M}{R^3})^{1/2}-1}
\end{align}
where we have substituted $w(R)$ from \eqref{I}. The above relation is the equation of motion of particles starting their radial free fall from rest at infinity in EGB space-time \eqref{I}. This is in accordance with Painlev\'e-Gullstrand coordinates used in \eqref{BB}. Introducing the scale factor $a( \tau)$ as a function of the proper time of the particles and substituting $ R ( \tau)=a( \tau) R_{c}$
for the surface of the star, one obtains
\begin{align}\label{U}
\big (\frac{\dot a}{a}\big)^2=\frac{1}{32\pi\alpha}\Big[\big(1+\frac{128\pi\alpha M}{a^3 R_c^3}\big)^{\frac{1}{2}}-1\Big]
\end{align}
which shows that 
we have restricted  ourselves to the spatially flat FRW interior geometry. \\
From \eqref{CC} one finds that the speed of collapse is proportional to $\tilde R^{1/4}$ for a sufficiently small radius. Therefore, similar to the regular BHs \cite{ourpaper}, the collapsing speed goes to zero at the center of the star. This is unlike OSD collapse where the collapsing star reaches the singularity of Schwarzschild BH with divergent speed. 
Equation \ref{CC} can be integrated to
\begin{align}
\frac{3}{8\sqrt{2 \pi \tilde \alpha}}\tilde\tau&=\bigg((1+\frac{64\pi\tilde{\alpha}}{\tilde{R}^3_{0}})^{1/2}-1\bigg)^{-1/2}\nonumber\\
&-\bigg((1+\frac{64\pi\tilde{\alpha}}{\tilde{R}^3})^{1/2}-1\bigg)^{-1/2}\nonumber \\
+& \frac{1}{\sqrt{2}}\Big(\tan^{-1}[\frac{1}{\sqrt{2}}(1+\frac{64\pi\tilde{\alpha}}{\tilde{R}^3})^{1/2}-1)^{1/2}]
\notag\\
&-\tan^{-1}[\frac{1}{\sqrt{2}}(1+\frac{64\pi\tilde{\alpha}}{\tilde{R}^3_0})^{1/2}-1)^{1/2}]\Big)\label{DD}
\end{align}
where the integration constant is chosen such that $\tilde{\tau}=0$ at $\tilde{R_0}$ and we have used the dimensionless variables defined in the previous section. For $\tilde{R}^3\gg64\pi \tilde{\alpha}$, this reduces to the corresponding one for OSD collapse, $3\tilde{\tau}/2=\tilde{R_0}^{3/2}-\tilde{R}^{3/2}$ \cite{blau}.
Expanding \eqref{DD} up to the first order in $\tilde R$ leads $\tilde R^{3/4} = - \beta \tilde \tau + \mathcal{G}(\tilde \alpha , \tilde R_0)$ in which $\mathcal{G}(\tilde \alpha , \tilde R_0)$ and $\beta$ are some constants. Therefore, at the late stages of collapse, the star's radius decreases slower than that of OSD model \cite{naresh,rez}. Setting $\tilde{R_0}=2$ and $\tilde{\alpha}=0.001$, the result \eqref{DD} is plotted in figure \ref{fig2}.  We see that, unlike the regular BHs \cite{ourpaper} with de Sitter cores such as Bardeen and Hayward BHs \cite{pnkl,lljo}, the collapse terminates after a finite time. \\
Moreover, considering \eqref{CC} and \eqref{DD}, one can obtain the evolution of Hubble parameter, $\tilde H( \tilde \tau)=\dot{ \tilde{R}}/\tilde{R}$ as following
\begin{align}
\frac{3}{2}\tilde{\tau}=\left(\frac{1}{\tilde{H}}-\frac{1}{\tilde{H}_0}\right)+4\sqrt{\pi\tilde{\alpha}}&\bigg(\tan^{-1}(4\tilde{H}_0 \sqrt{\pi\tilde{\alpha}})-\notag\\
&\tan^{-1}(4\tilde{H} \sqrt{\pi\tilde{\alpha}})\bigg)\label{FF}
\end{align}
in which $\tilde{H}_0$ is the initial value of Hubble parameter.
\section{Equation of state of collapsing star}
\label{sec4}
\begin{figure*}
\centering
\includegraphics[width=6.3in]{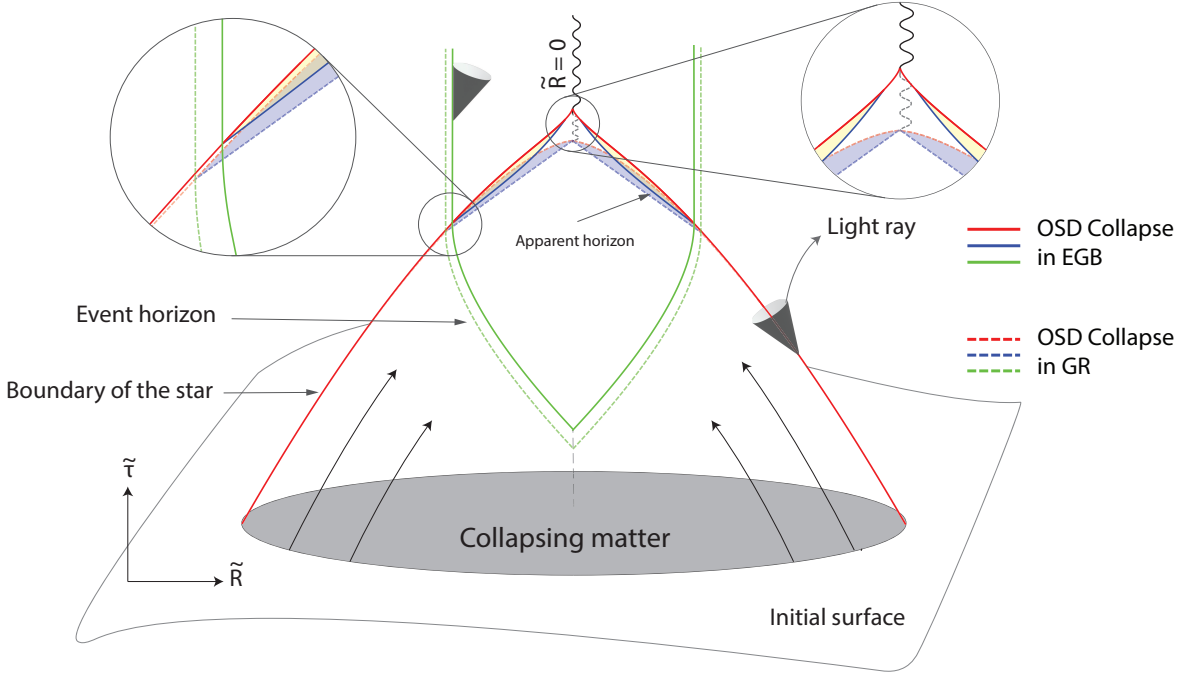}
\caption{The OSD collapse scenario in GR and 4D EGB gravity is shown by solid and dashed lines. Red, green, and blue lines describe the surface of star, the event horizon , and the apparent horizon respectively. The OSD collapse in 4D EGB gravity is plotted for $\tilde \alpha = 0.001$ and $\tilde{R}_0 = 2$. Such as OSD collapse, the end state is a finite time singularity.
}
\label{fig2}
\end{figure*}

Recalling that  the outside of star is not vacuum. Instead, it is filled with a perfect fluid whose source is in fact the GB term. Therefore, the interior geometry of star is governed by the standard Friedmann equations.
We show that the SEC is violated when the star radius
becomes smaller than a specific value.\\ 
We treat the high order curvature terms coming from 4D EGB gravity as an effective stress-energy tensor,
\begin{align}\label{PP}
 T_{\mu\nu}^{\text{eff}} & =  T_{\mu\nu}^{\text{EGB}} +  T_{\mu\nu}^{\text{star}} \hspace{0.5cm},\hspace{0.5cm}
 T_{\mu\nu}^{\text{EGB}} &=-2\alpha\mathcal{ H}_{\mu\nu}
\end{align}
where $T_{\mu\nu}^{\text{star}}$ presents the stress-energy tensor of collapsing ball of matter and we assume that each of the above tensors can be modeled by
anisotropic fluids.
Inserting \eqref{I} into the Einstein equations, one obtains
the corresponding energy density and anisotropic pressures of 4D EGB BH
as follows
\begin{align}
\frac{256\pi^2\tilde{\alpha}}{3}\tilde\rho^{\text{EGB}}&=-\frac{256\pi^2\tilde{\alpha}}{3}\tilde p^{\text{EGB}}_{r}\nonumber \\
&=\frac{\tilde r^3 + 32\pi \tilde \alpha}{\sqrt{\tilde{r}^3(\tilde{r}^3+64\pi\tilde{\alpha}})}-1\nonumber\\
\frac{256\pi^2\tilde{\alpha}}{3}\tilde p^{\text{EGB}}_{\theta}&= \frac{256\pi^2\tilde{\alpha}}{3}\tilde p^{\text{EGB}}_{\phi}\nonumber \\
&= 1-\frac{\tilde{r}^6 + 96\tilde{\alpha} \tilde {r}^3  +  512 \pi ^ 2 \tilde {\alpha} ^2}{\tilde {\alpha}\left (\tilde{r}\left(\tilde{r}^3+64\pi \tilde{\alpha}\right)\right )^{3/2}}
\label{WW}
\end{align}
Following \cite{ourpaper}, we assume that the geometry of star is described by spatially flat FRW metric, hence the effective stress-energy tensor of star is given by the perfect fluid one.  Moreover, the Friedmann equation \eqref{U}   would be identical with \eqref{CC} if
\begin{align}\label{QQ}
\tilde \rho^{\text{eff}}=\frac{3}{256 \pi^2  \tilde \alpha}\left
(\sqrt{1+\frac{64 \pi \tilde\alpha }{\tilde R^3}}-1\right)
\end{align}
It is clear that in the limit of $\tilde \alpha\to 0$, the density is proportional to $\tilde R^{-3}$ describing a dust fluid which is consistent with the OSD collapse. \\
Taking the time derivative of \eqref{QQ} and substituting it in the continuity equation, $\dot{\tilde \rho}^{\text{eff}}+3(\dot{\tilde R}/\tilde R)(\tilde\rho^{\text{eff}}+\tilde p^{\text{eff}})=0$, the stellar pressure would be obtained
\begin{align}\label{SS}
\frac{8 \pi}{3}\tilde{p}^{\text{eff}}= \frac{1}{32\pi \tilde\alpha} - \big( \frac{1}{\tilde R^3} + \frac{1}{32 \pi \tilde \alpha} \big)\big( 1 + \frac{64 \pi \tilde \alpha}{\tilde R^3}\big)^{-1/2}
\end{align}
\\During the collapse when $\tilde R\gg(64\pi\tilde\alpha)^{\frac{1}{3}}$, this equation reduces to $\tilde p^{\text{eff}}\simeq 0$ and the stellar matter would be pressureless. 
As mentioned above, the interior geometry of the star is assumed to be FRW geometry. Therefore, the stellar matter should have uniform density
and pressure given by \eqref{QQ} and \eqref{SS}. This is unlike the conventional OSD collapse for which the stellar matter is pressureless.
Now using \eqref{QQ} and \eqref{SS}, one can obtain  the stellar equation of state as \footnote{For
$\tilde \rho^{\text{eff}}\ll3/256\pi^2 \tilde\alpha$, \eqref{SSlp} can be expanded as
\begin{align}
\label{expandedp}
\tilde p^{\text{eff}}\simeq - \frac{128 \pi^2 \tilde \alpha (\tilde \rho^{\text{eff}})^2}{3} (1- \frac{256 \pi^2 \tilde \alpha}{3}\tilde \rho^{\text{eff}})
\end{align}
This is a special case of equation of state  
\begin{align}
\label{mboyne}
p_r = \bigg[ \zeta - (\zeta + 1 ) \bigg(\frac{\rho}{\rho^{\text{max}}}\bigg)^m \bigg] \bigg(\frac{\rho}{\rho^{\text{max}}}\bigg)^{1/n} \rho
\end{align}
propsed by \cite{mboyne} for a particular class of regular BHs with  de Sitter core. In \eqref{mboyne}, 
$\zeta$ is a free parameter, $n$ is the polytropic index and the core size is determined by $\rho^{\text{max}}$. The  anisotropic fluid \eqref{mboyne} with radial pressure $p_r$ and density $\rho$ is introduced as the source of such BHs \cite{mboyne}. Setting $ m =n =1$,  
$\zeta = - \rho^{\text{max}}/4 \pm \sqrt{(\rho^{\text{max}}/4)^2 - \rho^{\text{max}}/2}$
and $\rho^{\text{max}}= 3/ 256 \pi^2 \tilde \alpha $,  equation \eqref{expandedp} is now the same as \eqref{mboyne}.}
\begin{align}\label{SSlp}
\tilde p^{\text{eff}}=-\frac{128 \pi^2 \tilde \alpha  (\tilde \rho^{\text{eff}})^2}{3} (1+ \frac{256 \pi^2 \tilde \alpha}{3}\tilde \rho^{\text{eff}})^{-1}
\end{align}
It is clear that for sufficiently low energy densities, equation \eqref{SSlp} can be rewritten as $\tilde p^{\text{eff}}=-128\pi^2 \tilde \alpha (\tilde\rho^{\text{eff}})^2/3$. It is the equation of state of a polytropic fluid with the unit polytropic index, obtained in \cite{ourpaper} for OSD collapse in Hayward space-time.
On the other hand, for sufficiently large density, \eqref{SSlp} reduces to the linear equation of state $\tilde p^{\text{eff}}=-\tilde \rho^{\text{eff}}/2$.\\  
The presence of a negative sign in \eqref{SSlp} indicates that some energy conditions may be violated. A simple calculation, taking into account \eqref{QQ} and \eqref{SS}, shows that all energy conditions are satisfied except the SEC,
$\tilde \rho^{\text{eff}}+3\tilde p^{\text{eff}}\le 0$, which is violated for
\begin{align}\label{VV}
 \tilde{R}\le (8 \pi\tilde{\alpha})^{\frac{1}{3}}
\end{align}
Therefore, the SEC is violated when the star radius
becomes smaller than a specific value given by \eqref{VV} and this slows the rate of contraction at late
times\footnote{There exists a similar inequality when a regular BH with an asymptotically de Sitter core, is formed from the OSD gravitational collapse of a massive star \cite{ourpaper}. In that case, there is a free parameter in the space-time metric which determines the core radius. This parameter also appears on the right-hand side of the mentioned inequality and prevents the formation of singularity.} \cite{naresh}, see figure \ref{fig2}. The violation of SEC is the result of applying a smooth transition from a collapsing spatially flat FRW geometry to the 4D EGB BH.
However, the repulsive effects can not prevent
the formation of singularity at the center of collapsing star. 
The star evolves to the singularity at a finite
time like OSD collapse. 
\\
As mentioned before, the relations \eqref{QQ} and \eqref{SS} represent
the total energy density and pressure on the surface of
star and its interior according to \eqref{PP}.
Taking into account \eqref{WW}, one can obtain the pure contribution of star pressure 
to the total pressure as $\tilde p^{\text{star}}=\tilde p^{\text{eff}}-\tilde p^{\text{EGB}}\big|_{\tilde r=\tilde R}=0$. 
Therefore, as we have expected, the pure radial pressure on the surface of the star is zero.\\
Let us now investigate the gravitational mass of collapsing star in 4D EGB space-time accepting the first approach. 
Equation \eqref{QQ} gives the following mass function
\begin{align}\label{AAA}
M&=
\frac{4}{3}\pi R^3 \rho^{\text{eff}}\left(1+\frac{128 \pi^2 \alpha}{3}\rho^{\text{eff}}\right)
\end{align}
The total mass of the star is $M_{s}=\frac{4}{3}\pi R^3\rho^{\text{eff}}$, therefore
\begin{align}\label{BBB}
\frac{M }{M_{s}} &= 1 + \frac{32\pi \alpha}{ R^3} M_s
\end{align}
As expected, for $\alpha\to 0$ the total mass of the star and its gravitational mass are the same.
\section{Horizons}
\label{sec5}
\begin{figure}[h]
\centering
\includegraphics[width=3.45in]{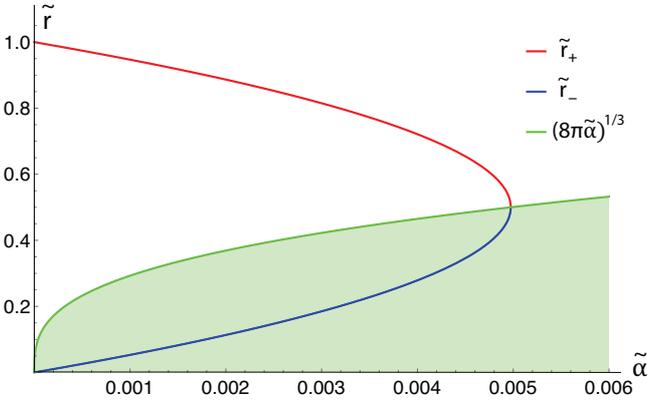}
\caption{$\tilde{r}_+$ and $\tilde{r}_-$ with respect to $\tilde{\alpha}$. The SEC is violated in the shaded region.} 
\label{fig1}
\end{figure}

In this section, we will study the properties of apparent and event horizons. Outside the collapsing object, the event and apparent horizons coincide and can be found from \eqref{I} \cite{plp}
\begin{equation}\label{HH}
\tilde r^2-\tilde r+16\pi\tilde\alpha=0
\end{equation}
This equation has two real positive roots for $\tilde{\alpha}\le1/64\pi$,
\begin{align}\label{II}
\tilde{r}_{\pm}=\frac{1}{2}\big(1\pm\sqrt{1-64\pi\tilde{\alpha}}\big)
\end{align}
where $\tilde{r}_{+}$ and $\tilde{r}_{-}$ are outer and inner horizons of the BH.
For $\tilde\alpha = 1/64\pi$, we have an extremal BH and for $\tilde\alpha >1/64\pi$ we have a horizonless compact object. The locations of horizons and the region of violation of SEC \eqref{VV}, is plotted in figure \ref{fig1} with respect to $\tilde{\alpha}$. 
The upper and lower curves represent $\tilde{r}_{+}$  and $\tilde{r}_{-}$  respectively.
\\Assuming that at $\tilde{\tau}_{f}$, the radius of the star crosses the event horizon $\tilde{r}_{+}$, i.e. $\tilde{R}(\tilde{\tau}_{f})=\tilde{r}_{+}$. From \eqref{DD}, we will have 
\begin{align}
\frac{3}{8\sqrt{2\pi\tilde{\alpha}}}\tilde\tau_{f}&=\bigg[ \big((1+\frac{64\pi\tilde{\alpha}}{\tilde{R}^3_{0}})^{1/2}-1\big)^{-1/2}-\frac{\tilde{r}_{+}}{4\sqrt{2\pi\tilde{\alpha}}}\bigg]+\nonumber \\
\frac{1}{\sqrt{2}}\Bigg(\tan^{-1}&\frac{4\sqrt{\pi\tilde{\alpha}}}{\tilde{r}_{+}}-
\tan^{-1} \frac{1}{\sqrt{2}}\sqrt{(1+\frac{64\pi\tilde{\alpha}}{\tilde{R}^3_{0}})^{\frac{1}{2}}-1}\Bigg)\label{JJ}
\end{align}
\\in which $(1+64\pi\tilde{\alpha}/\tilde{r}^3_+)^{1/2}-1=32\pi\tilde{\alpha}/\tilde{r}^2_+$ is used according to \eqref{HH}.\\
In the interior of star,  the radial null geodesics are given by $\dot{\tilde r}=\pm1+\tilde r\tilde H$ using \eqref{EE}. Since $\tilde H<0$, the negative sign corresponds to the ingoing null geodesics while the positive sign represents ingoing and outgoing null geodesics for $\tilde r>-1/\tilde H$ and
$\tilde r<-1/\tilde H$ respectively.
Therefore, the trapped surfaces are inside $\tilde r_{ah}=-1/\tilde H$, which is the apparent horizon, and thus a BH is formed. To find the evolution of the apparent horizon, we write \eqref{FF} as a function of $\tilde{r}_{ah}$,
\begin{align}
\frac{3}{2}\tilde{\tau}=-\left(\tilde{r}_{ah}+\frac{1}{\tilde{H}_0}\right)+ 4\sqrt{\pi\tilde{\alpha}}&\bigg(\tan^{-1}(4 \sqrt{\pi\tilde{\alpha}}\tilde{H}_0)+\notag\\
&\tan^{-1}(\frac{ 4\sqrt{\pi\tilde{\alpha}}}{\tilde{r}_{ah}})\bigg)\label{KK}
\end{align} 
where the initial radius of apparent horizon is chosen  to be $-1/\tilde{H}_0$ at $\tilde{\tau}=0$ (figure \ref{fig2}). It is easy to see that 
$\tilde{r}_{ah}(\tilde{\tau}_{f})=\tilde r_{+}$, as expected.
Using \eqref{KK} and calculating the metric \eqref{EE} on the apparent horizon, one can obtain the norm of the normal vector of the apparent horizon as follows
\begin{align}
n_{\alpha}n^{\alpha}=\frac{3}{4}\big ( \frac{ \tilde r_{\text{ah}}^2 + 16 \pi \tilde \alpha + 16 \sqrt{ \pi \tilde \alpha}}{\tilde r_{\text{ah}}^2+16\pi \tilde\alpha}\big)
\end{align}
Therefore, the apparent horizon remains timelike during the collapse.\\
In the following, we are going to investigate the evolution of the interior event horizon, $\tilde{r}_{eh}$. To do this, we should consider the outgoing null geodesic arriving the surface of star at $\tilde{\tau}_{f}$ for which $\tilde{R}(\tilde{\tau}_{f})=\tilde{r}_{+}$. From \eqref{EE}, the outgoing null geodesics could be written as $\dot{\tilde{r}}=1+\tilde{r}\tilde{H}$.
This leads to the following equation for the event horizon using $\dot{\tilde{r}}=(d\tilde{r}/d\tilde{R})\dot{\tilde{R}}$  and expression \eqref{CC} for the Hubble parameter
\begin{equation}\label{MM}
\frac{d\tilde{r}_{eh}}{d\tilde{R}}-\frac{\tilde{r}_{eh}}{\tilde{R}}+\frac{4\sqrt{2\pi\tilde{\alpha}}}{\tilde{R}}\big( (1+64\pi\tilde{\alpha}/\tilde{R^3})^{1/2}-1\big)^{-1}=0
\end{equation}
which gives $\tilde{r}_{eh}$ as a function of $\tilde{R}$
\begin{align}\label{NN}
\tilde{r}_{eh}=&\tilde{R}+\frac{2^{\frac{17}{6}}3}{5}\frac{\sqrt{\pi\tilde{\alpha}}\sqrt{x-1}}{(x+1)^{\frac{1}{3}}}{_{2} F_ 1\left (\frac {2} {3}, \frac {5} {6};\frac {11} {6};\frac {1-x} {2} \right)}\notag\\& -\frac{\tilde{R}}{\tilde{r}_{+}}\frac{2^{\frac{17}{6}}3}{5}\frac{\sqrt{\pi\tilde{\alpha}}\sqrt{x_{+}-1}}{(x_{+}+1)^{\frac{1}{3}}}{_{2} F_ 1\left (\frac {2} {3}, \frac {5} {6};\frac {11} {6};\frac {1-x_{+}} {2} \right)}\notag\\&-\frac{8\sqrt{2\pi\tilde{\alpha}}}{\sqrt{x-1}}+
\frac{\tilde{R}}{\tilde{r}_{+}}\frac{8\sqrt{2\pi\tilde{\alpha}}}{\sqrt{x_{+}-1}}
\end{align}
where $x=(1+64\pi\tilde{\alpha}/\tilde{R^3})^{1/2}$ and $x_{+}=(1+64\pi\tilde{\alpha}/\tilde{r}_{+}^3)^{1/2}$. \\
Considering \eqref{DD}, \eqref{KK} and \eqref{NN}, the basic features of OSD collapse to 4D EGB BH are plotted in figure \ref{fig2} setting $\alpha=0.001$ and $\tilde{R}_0 = 2$.
We see that, unlike OSD collapse, when the star contracts sufficiently, the trapped surfaces may be disappeared. Moreover, the star surface in our model evolves slower than OSD collapse. The same happens for event and apparent horizons. 
These prperties also hold for a regular BH \cite{ourpaper}. However, here such as OSD scenario, the gravitational collapse is ended in a finite time, therefore the singularity is not avoided \cite{naresh}.\\
It should be noted that in 4D EGB BH, the gravitational force is repulsive at a sufficiently small radius \cite{plp} while the curvature scalar diverges as $R\propto r^{-\frac{3}{2}}$.
At the first sight it may be expected that after horizon formation, the gravitational collapse stops when the repulsive effects come into.
As shown here,
it takes a finite proper time for the star to contract completely,  such as the OSD collapse in GR. However, for the 4D EGB BH, as it is shown in figure \ref{fig2}, the star radius decreases slower and therefore, it takes more time to reach the singularity \cite{naresh}. 
\section{Concluding Remarks}\label{concluding}
In this paper, we have investigated the gravitational collapse of a massive object to the 4D EGB BH in the framework of the OSD model. The inside geometry is described by spatially flat FRW metric which is joined smoothly to the EGB BH space-time. To describe the stellar matter, we consider that the cosmological dynamics is governed by the standard FRW equations and outside the star is not vacuum. The density and pressure of the star are determined by the smooth matching of geometries at the surface of the star resulted from GR junction conditions. \\
According to \eqref{DD}, we have shown that after a finite time, the star radius goes to zero, the energy density \eqref{QQ} becomes infinite and thus the curvature scalar diverges. However, in this limit the BH metric \eqref{I} remains finite
while the gravitational force is repulsive
\cite{plp} and behaves as $r^{-1/2}$. According to the modified OSD gravitational collapse presented here,
the collapsing matter reaches the singularity with zero velocity finally. Therefore, the speed of collapse in the 4D EGB case is slower than OSD collapse \cite{naresh}.\\
It is useful to remember that for the regular BHs \cite{ourpaper}, the curvature scalar and metric remain finite in the limit of zero radius and the OSD collapse of the star is ended in an infinite time. However for the 4D EGB spherically symmetric geometry, the metric approaches a finite value  at the mentioned limit whereas the curvature scalar diverges and a finite time is needed to complete the collapse. Now a question may be raised here. Is there any relation between the final collapsing speed of the star and the regularity of the exterior metric function in the limit $r\rightarrow 0$? To answer this question, let us consider the metric \eqref{I} without specifying $\omega(r)$. Then from \eqref{CC}, one can easily see that the collapsing speed is proportional to $(r\omega)^{-1/2}$. Thus, in the limit $r\rightarrow 0$,  the final collapsing speed is infinite if and only if the metric function is singular.
\\We have obtained the dynamics of the interior apparent and event horizons and also the stellar surface as functions of the proper time of star. It is shown that such as OSD model, at the first stage of collapse, the event horizon starts to increase from a zero radius. Then, when the star surface intersects the event horizon, the apparent horizon starts to decrease from $r_+$. The rate of evolution of these surfaces depends on the value of $\alpha$.\\

\textbf{Acknowledgments:}
F. Shojai is grateful to the
University of Tehran for supporting this work under a
grant provided by the university research council.


\begin{thebibliography}{99}
\bibitem{klsh}
R. Penrose, Phys. Rev. Lett. \textbf{14}, 57–59 (1965);
Hawking, S.W.; Penrose, R. , Proc. R. Soc. Lond.
A, 314, 529–548 (1970); Hawking, S.W. , Phys.
Rev. D, \textbf{14}, 2460–2473 (1976).


\bibitem{quantum}
 V.P. Frolov, G.A. Vilkovisky, In Proceedings of the Second Marcel Grossmann Meeting on General Relativity, Trieste, Italy, 5–11 July (1979); V.P. Frolov, G.A. Vilkovisky, Phys. Lett. B, \textbf{106}, 307–313 (1981).

\bibitem{dhj}
E. Ayon-Beato,  A. Garcia, Phys. Rev. Lett. \textbf{80}, 5056  (1998).

\bibitem{pnkl}
S. A. Hayward, Phys. Rev. Lett. \textbf{96}, 031103 (2006).

\bibitem{lljo}
J.M. Bardeen, in proceedings of the International Conference GR5, Tbilisi, U.S.S.R. (1968).

\bibitem{sakh}
 A.D. Sakharov, Sov. Phys. JETP \textbf{22} 241 (1966).

\bibitem{45-46Mal}
N. Itoh, Prog. Theor. Phys. \textbf{44}, 291–292 (1970);
E. Witten, Phys. Rev. D \textbf{30}, 272–285 (1984)

\bibitem{47-48Mal}
R. Ruffini, S. Bonazzola, Phys. Rev. \textbf{187}, 1767–1783 (1969);
F.E. Schunck, and E.W. Mielke, Class. Quantum Grav. \textbf{20}, R301–R356 (2003).

\bibitem{34-35Mal}
 V.P. Frolov, G.A. Vilkovisky, In Proceedings of the Second Marcel Grossmann Meeting on General Relativity, Trieste, Italy, 5–11 July (1979); V.P. Frolov, G.A. Vilkovisky, Phys. Lett. B, \textbf{106}, 307–313 (1981).

\bibitem{25Mal}
P.~O.~Mazur and E.~Mottola,
 arXiv:gr-qc/0109035.

\bibitem{44Mal}
C. Barceló, S. Liberati, S. Sonego, M. Visser, Phys. Rev. D, \textbf{77}, 044032 (2008).

\bibitem{ourpaper}
F.~Shojai, A.~Sadeghi and R.~Hassannejad,
Class. Quantum Grav. \textbf{39}, 8, 085003 (2022).

\bibitem{OSD}
J. R. Oppenheimer and H. Snyder, Phys. Rev. \textbf{56}, 455 (1939); S. Datt, Zs. f. Phys. \textbf{108}, 314 (1938).



\bibitem{plp}
D.~Glavan and C.~Lin,
Phys. Rev. Lett. \textbf{124}, 8, 081301  (2020).




























\bibitem{nilmk}
P.~G.~S.~Fernandes, P.~Carrilho, T.~Clifton and D.~J.~Mulryne,
Phys. Rev. D \textbf{102}, 2, 024025 (2020)

\bibitem{ppgf}
 J. Bonifacio, K. Hinterbichler, and L. A. Johnson, Phys. Rev, D, 102(2), 024029 (2020).

\bibitem{ppkk}
 T.~Kobayashi,
JCAP \textbf{07} , 013 (2020).


\bibitem{nkb}
L. Ma, and H. Lü, Eur. Phys. J. C, 80(12), 1-10 (2020).

\bibitem{hbjb}
S.~L.~Li, P.~Wu and H.~Yu,
arXiv:2004.02080 [gr-qc] (2020).

\bibitem{jnk}
M.~A.~Garc\'\i{}a-Aspeitia and A.~Hern\'andez-Almada,
Phys. Dark Univ. \textbf{32}, 100799 (2021).

\bibitem{vfsk}

R. A. Konoplya, and A. Zhidenko, Phys. Rev. D, \textbf{101}, 08438 (2020).

\bibitem{njj}
 R. A. Hennigar, D. Kubizňák, R. B. Mann, and C. Pollack, Phys. Lett. B, \textbf{808}, 135657  (2020).


\bibitem{Kumar}
R.~Kumar and S.~G.~Ghosh,
JCAP \textbf{07}, 053 (2020).

\bibitem{klkl}
P.~G.~S.~Fernandes,
Phys. Lett. B \textbf{805}, 135468 (2020).

\bibitem{xzsrz}
K. Yang, B. M. Gu, S. W. Wei, and  Y. X. Liu, Eur. Phys. J. C, \textbf{80}(7), 662 (2020).

\bibitem{pkjk}
M. Guo, and P. C.  Li,  Eur. Phys. J. C, \textbf{80}(6), 588 (2020).

\bibitem{vfyj}
 R. A. Konoplya, and A. F. Zinhailo,  Eur. Phys. J. C, \textbf{80}, 1049 (2020).

\bibitem{xtcy}
X. X. Zeng, H. Q. Zhang, and H. Zhang, Eur. Phys. J. C, \textbf{80}, 872  (2020).

\bibitem{oorz}
R.~Roy and S.~Chakrabarti,
Phys. Rev. D, \textbf{102}(2), 024059 (2020).

\bibitem{xffy}
M.~S.~Churilova,
Phys. Dark Univ. \textbf{31}, 100748 (2021)

\bibitem{agji}
A.~Arag\'on, R.~B\'ecar, P.~A.~Gonz\'alez and Y.~V\'asquez,
Eur. Phys. J. C \textbf{80}(8), 773 (2020).

\bibitem{pvdt}
S. W. Wei, and  Y. X. Liu,  Phys. Rev. D, \textbf{101}, 104018 (2020).

\bibitem{gfyu}
D. V.Singh, and S. Siwach, Phys. Lett. B, \textbf{808}, 135658 (2020).

\bibitem{spsjj}
S. A. H. Mansoori, Phys. Dark Univ. \textbf{31}, 100776 (2021).

\bibitem{dncdkl}
S. Ying, Chinese Phys. C, \textbf{44}, 125101 (2020).

\bibitem{aeth}
C. Y. Zhang, P. C. Li, and M. Guo, Eur. Phys. J. C, \textbf{80}(9), 890  (2020).

\bibitem{cfdt}
 R. A. Konoplya, and A. F. Zinhailo, Phys. Lett. B, \textbf{810}, 135793 (2020).

\bibitem{jjkrz}
K. Aoki, M. A. Gorji, and S. Mukohyama, JCAP \textbf{09}, 014 (2020).



\bibitem{njrf}
 S. I. Nojiri, and S. D. Odintsov, EPL (Europhysics Letters), \textbf{130}(1), 10004 (2020).

\bibitem{nsxd}
 D. Wang, and D. Mota, Phys. Dark Univ. \textbf{32}, 100813 (2021).

\bibitem{arre}
J.~Arrechea, A.~Delhom and A.~Jim\'enez-Cano,
Phys. Rev. Lett. \textbf{125}, 14, 149002  (2020).

\bibitem{aoki}
K.~Aoki, M.~A.~Gorji and S.~Mukohyama,
Phys. Lett. B \textbf{810}, 135843  (2020);  K.~Aoki, M.~A.~Gorji, S.~Mizuno and S.~Mukohyama,
JCAP \textbf{01}, 054  (2021).

\bibitem{jafarzade}
K.~Jafarzade, M.~Kord Zangeneh and F.~S.~N.~Lobo,
JCAP \textbf{04}, 008 (2021).

\bibitem{fernandes1}
R.~A.~Hennigar, D.~Kubiz\v{n}\'ak, R.~B.~Mann and C.~Pollack,
JHEP \textbf{07}, 027 (2020).

\bibitem{lu}
H.~Lu and Y.~Pang,
Phys. Lett. B \textbf{809}, 135717 (2020).

\bibitem{fer7}
P.~G.~S.~Fernandes, P.~Carrilho, T.~Clifton and D.~J.~Mulryne,
Class. Quant. Grav. \textbf{39}, 6, 063001 (2022).

\bibitem{fer8}
P.~G.~S.~Fernandes, P.~Carrilho, T.~Clifton and D.~J.~Mulryne,
Phys. Rev. D \textbf{104}, no.4, 044029  (2021).

\bibitem{naresh}
D.~Malafarina, B.~Toshmatov and N.~Dadhich,
Phys. Dark Univ. \textbf{30}, 100598 (2020).

\bibitem{kokk}
D. Lovelock, Journal of Mathematical Physics, \textbf{12}, 498-501 (1971).



\bibitem{f(r)}
S. I. Nojiri, and S. D. Odintsov, Int. J. Geom. Methods Mod. Phys. \textbf{4}(01), 115-145  (2007); 
F. S. N. Lobo, M. A. Oliveira, Phys. Rev. D \textbf{80}, 104012 (2009).

\bibitem{curvmatt}
T. Harko,  F.S.N. Lobo, Galaxies, \textbf{2}(3), 410-465 (2014). 
N. M. Garcia, F. S. N. Lobo, Phys. Rev. D \textbf{82}, 104018 (2010).

\bibitem{manheim}
P. D. Mannheim, Progress in Particle and Nuclear Physics, \textbf{56}(2) , 340-445 (2006).
F. S. N. Lobo, Class. Quantum Grav. \textbf{25}, 175006 (2008).
 
\bibitem{sepangi}
S. Shahidi, and H. R. Sepangi, Int. J.  Mod. Phys. D, \textbf{20}(01), 77-91 (2011).
F. S. N. Lobo, Phys. Rev. D \textbf{75}, 064027 (2007).

\bibitem{boulware}
D. G. Boulware, and S. Deser Phys. Rev. Lett. \textbf{55}, 2656 (1985).

\bibitem{saha}
C. Sahabandu, P. Suranyi, C. Vaz, and L. C. R. Wijewardhana Phys. Rev. D \textbf{73}, 044009 (2006).





\bibitem{kojfg}
W. Israel, Nuovo Cimento, \textbf{44}, 4349 (1966).

\bibitem{klpk}
P. Painlevé, C. R. Acad. Sci. \textbf{173}, 677–680 (1921).
A. Gullstrand, Ark. Mat. Astr. Fys. \textbf{16}(8): 1–15 (1922).
K. Martel and E. Poisson, Am. J. of Phys. \textbf{69}, 476-480 (2001).


\bibitem{dev}
S.~C.~Davis,
Phys. Rev. D \textbf{67} (2003), 024030

\bibitem{blau}
M.~Blau, "Lecture note on general Relativity",  http://www.blau.itp.unibe.ch/newlecturesGR.pdf (2020)

\bibitem{rez}
L. Rezzolla,  "An Introduction to Gravitational Collapse to Black Holes", Lectures given at the Villa Mondragone International School of Gravitation and Cosmology, Sept. 7th – 10th, 2004 Frascati (Rome), Italy

\bibitem{mboyne}
M. R. Mbonye and D. Kazanas, Phys. Rev. D \textbf{72}, 024016 (2005).
%














\end{thebibliography}
\end{document}